\documentclass[reprint,twocolumn,superscriptaddress,amsmath,amssymb,showpacs,prl]{revtex4-1}
\usepackage{hyperref}
\usepackage{color}
\pdfoutput=1 

\usepackage{dcolumn}
\usepackage{epsfig}
\usepackage{graphics}
\usepackage[latin1]{inputenc} 
\usepackage[T1]{fontenc}
\usepackage{bm}
\usepackage{hyperref}
\newcommand{\ddst}{false}

\begin{document}

\title{Nano-Ductility in Silicate Glasses is Driven by Topological Heterogeneity}

 \author{Bu Wang}
 \affiliation{Physics of AmoRphous and Inorganic Solids Laboratory (PARISlab), Department of Civil and Environmental Engineering, University of California, Los Angeles, CA, USA}
 \author{Yingtian Yu}
 \affiliation{Physics of AmoRphous and Inorganic Solids Laboratory (PARISlab), Department of Civil and Environmental Engineering, University of California, Los Angeles, CA, USA}
 \author{Mengyi Wang}
 \affiliation{Physics of AmoRphous and Inorganic Solids Laboratory (PARISlab), Department of Civil and Environmental Engineering, University of California, Los Angeles, CA, USA}
 \author{John C. Mauro}
 \affiliation{Science and Technology Division, Corning Incorporated, Corning, New York 14831, USA}
\author{Mathieu Bauchy}
 \email[Contact: ]{bauchy@ucla.edu}
 \homepage[\\Homepage: ]{http://mathieu.bauchy.com}
 \affiliation{Physics of AmoRphous and Inorganic Solids Laboratory (PARISlab), Department of Civil and Environmental Engineering, University of California, Los Angeles, CA, USA}
 
\date{\today}

\pacs{61.43.-j, 62.40.+i, 62.25.Mn}

\begin{abstract}
The existence of nanoscale ductility during the fracture of silicate glasses remains controversial. Here, based on molecular dynamics simulations coupled with topological constraint theory, we show that nano-ductility arises from the spatial heterogeneity of the atomic network's rigidity. Specifically, we report that localized floppy modes of deformation in under-constrained regions of the glass enable plastic deformations of the network, resulting in permanent change in bond configurations. Ultimately, these heterogeneous plastic events percolate, thereby resulting in a non-brittle mode of fracture. This suggests that nano-ductility is intrinsic to multi-component silicate glasses having nanoscale heterogeneities.
\end{abstract}

\maketitle

Although silicate glasses are commonly viewed as archetypal brittle materials, the existence of metal-like ductility at the nanoscale has recently been suggested \cite{Celarie2003, Chen2007, Shi2014, Wang2015}. This has both fundamental and practical importance, as increasing such ductility would allow one to design tougher glasses. Such glasses, more resistant to fracture while retaining their transparency, would broadly expand the range of applications for glasses \cite{Mauro2014}.

However, the existence of nano-ductility in glass remains highly debated. Celarie first reported the observation of nano-ductility in an aluminosilicate glass via fractured surface topographical analysis \cite{Celarie2003}. However, a later study using atomic force microscopy mapping the fractured surfaces of silica and soda-lime glass did not find any evidence of such a ductile failure \cite{Guin2004} and neither did cathodoluminescence spectroscopy measurements on silica \cite{Pezzotti2009}. In an effort to resolve this debate, simulations have also been conducted to explore the relationship between ductility and fine structural details, e.g. nano-cavities \cite{Chen2007}, or material properties, e.g. the Poisson's ratio \cite{Shi2014}. In particular, our recent study showed that the composition of glass plays a critical role in determining the existence of nano-ductility \cite{Wang2015}. Such a compositional dependence, which might have partly contributed to the discrepancies among experiments, is in agreement with experimental results obtained for sodium silicate glasses \cite{Bonamy2006}.

Besides the extrinsic origins theorized in previous studies, e.g., \textcolor{red}{stress corrosion cracking with the presence of water} \cite{Guin2004,Lechenault2011a,Lechenault2011b,Smedskjaer2015} or macroscopic defects \cite{Chen2007}, questions remain regarding the atomistic origin of such nano-ductility for suitable compositions of glass. In particular, spectroscopy analyses revealed the existence of an excess of alkali network modifier atoms in as-fractured surfaces of silicate glass \cite{Almeida2014}, which suggests that crack propagation may preferentially occur along alkali-rich regions, in agreement with the picture offered by Greaves' modified random network \cite{Greaves1985}. Here, based on molecular dynamics (MD) simulations coupled with topological constraint theory \cite{Mauro2011, Phillips1979, Phillips1981}, we show that such spatial fluctuations of alkali's concentration induce heterogeneities in the network's local rigidity \cite{Bauchy2013}, which, in turn, can result in crack deflection \cite{Lawn1994} and ductility. This suggests that nano-ductility is intrinsic to multi-component silicate glasses and originates from topological heterogeneities.

Topological constraint theory has proven to be a powerful tool to evaluate the important rigidity of atomic networks, while filtering out the chemical details that ultimately do not affect macroscopic properties \cite{Mauro2011, Phillips1979, Phillips1981}. As such, it has provided critical understanding of the atomic origins of various phenomena in glasses, such as fracture statistics and composition-dependent hardness \cite{Bauchy2015, Mauro2012, Smedskjaer2014, Smedskjaer2010, Bauchy2014}. Analogously to mechanical trusses, the rigidity of an atomic network can be evaluated by enumerating the number of constraints per atom ($n_{\rm c}$), which includes bond-stretching and bond-bending constraints, and comparing this metrics with the number of degrees of freedom per atom (3 for three-dimensional networks). Under-constrained structures ($n_{\rm c} < 3$, flexible) contain extra internal degrees of freedom ($f = 3 - n_{\rm c}$, floppy modes \cite{Naumis2005}) and, thereby, feature low-energy modes of deformation \cite{Cai1989}, which allow flexibility in atomic rearrangement and structural relaxation. In contrast, over-constrained structures ($n_{\rm c} > 3$, stressed-rigid) become rigid and undergo internal eigen-stress \cite{Wang2005}. In between, the existence of an isostatic intermediate phase has also been suggested \cite{Boolchand2001}, in which networks are rigid but free of eigen-stress \cite{Micoulaut2007} or consist of a combination of rigid and floppy regions \cite{Chubynsky2006}. Such isostatic glasses have been shown to feature maximal fracture toughness \cite{Varshneya2007}, which suggests that the resistance to fracture is related to the atomic topology \cite{Bauchy2014}.

In our previous study, we reported that pure silica breaks in a nearly perfectly brittle manner, while the fracture of multi-component glasses such as sodium silicate and calcium aluminosilicate exhibit significant ductility \cite{Wang2015}. \textcolor{red}{To understand the atomistic origin of this ductility, we simulate the fracture of (Na$_2$O)$_{20}$(SiO$_2$)$_{80}$ glasses made of 9000 atoms (hereafter denoted as 20NS) following a procedure previously described \cite{Wang2015}. The glass structures are obtained by melting random atomic configurations at 4000 K for 1ns and then quenching the glass-forming liquids to 300 K with 1 K/ps cooling rate, all in NPT ensemble with zero pressure.} After an equilibration at 300 K for 1 ns under zero pressure, the simulation box is gradually stretched by stepwise 0.5\% \textcolor{red}{($\sim$ 0.25 \AA)} increases along the $z$ direction, until the structure is fully fractured. \textcolor{red}{During each step, the structure is first stretched by linearly scaling the atomic coordinates. The system is then relaxed for 50 ps before a statistical averaging phase of 50 ps, all in the NVT ensemble. Note that mimicking standard tests of fracture toughness $K_{Ic}$ would require to initially insert a notch in the simulation box \cite{Wang2015}. However, here, rather than measuring $K_{Ic}$, we aim to observe the spontaneous global response of the system to a tensile stress. As such, no notched is inserted here, as it would arbitrarily concentrate the stress in a pre-determined region of the glass.}

\begin{figure}
\begin{center}
\includegraphics*[width=\linewidth, keepaspectratio=true, draft=\ddst]{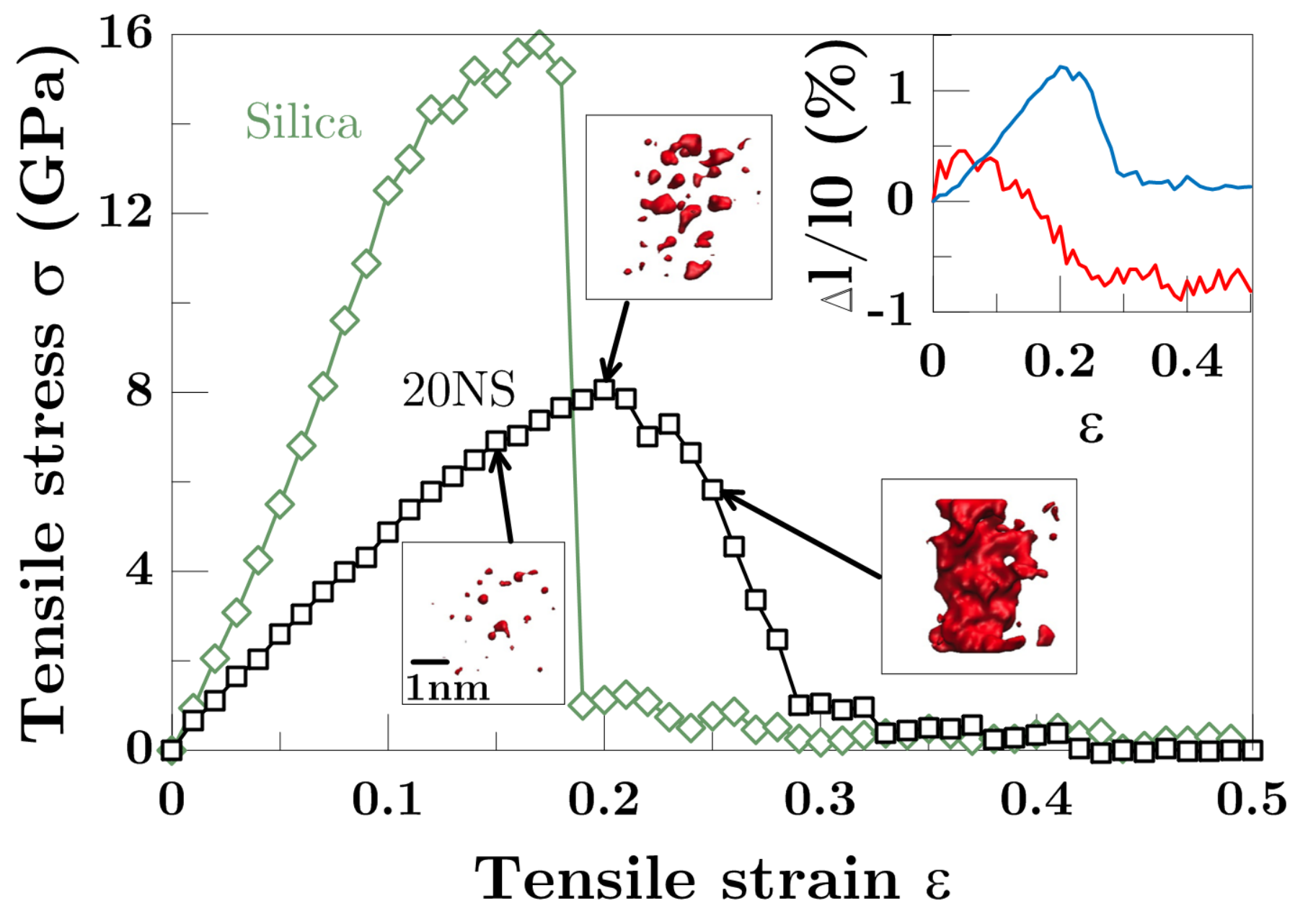}
\caption{\label{fig:ss} (Color online) Stress-strain response of the simulated (Na$_2$O)$_{20}$(SiO$_2$)$_{80}$ glass during fracture, \textcolor{red}{compared with that of pure silica}. The snapshots illustrate the volume of the cavities, with a radius larger than 5 \AA, that are formed at various strains. The inset shows the relative variation of the average Si--O and Na--O bond lengths ($\Delta l/l_0$) with respect to the strain. The cutoffs used to identify Si--O and Na--O bonds are 1.974 \AA\  and 3.311 \AA, respectively, as determined from the position of the first minimum after the peak associated to the first coordination shell in the pair distribution functions.
}
\end{center}	
\end{figure}

Figure \ref{fig:ss} shows the computed tensile stress with respect to the applied strain during the fracture. In agreement to what was observed for 30NS \cite{Wang2015}, 20NS exhibits a non-brittle fracture behavior, i.e., the stress does not suddenly decrease to zero after reaching its maximum, when the crack starts to propagate. \textcolor{red}{Such behavior strongly contrasts with that observed for the fracture of pure silica, in which a sudden drop of stress is observed (see Figure \ref{fig:ss})}. The simulation reveals the existence of cavities that form during the fracture, as typically observed for ductile materials. Indeed, as shown in Fig. \ref{fig:ss} for a series of strains, the fracture clearly happens through cavity initiation, growth and, eventually, coalescence. \textcolor{red}{This mechanism is in agreement with previous simulations, which report that nano-ductility arises from pre-existing, nano-sized voids in pure silica \cite{Chen2007}. Here, in the case of sodium silicate, we show that the formation of such cavities can naturally happen, even without pre-existing microscopic defects.} We note, however, that one should be cautious about using such phenomenological observation in experiment to qualify the nature of the fracture, especially from surface measurements. Indeed, as cavities formation and crack propagation occur in the bulk volume, it would be challenging to distinguish newly formed voids, appearing in front of the crack tip, from cracks propagating perpendicular to the surface.

\begin{figure}
\begin{center}
\includegraphics*[width=\linewidth, keepaspectratio=true, draft=\ddst]{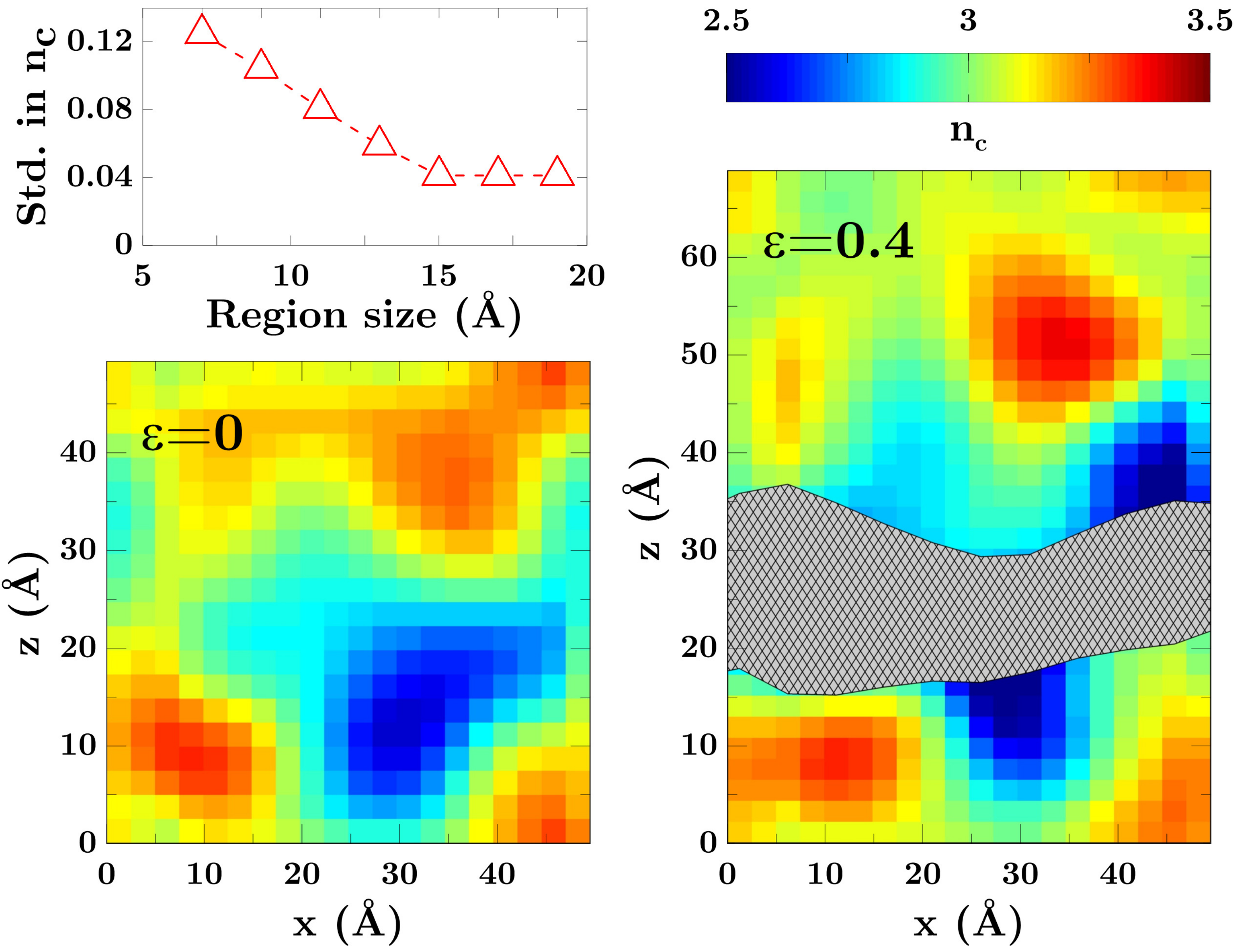}
\caption{\label{fig:map} (Color online) Contour maps of the local number of constraints per atom ($n_{\rm c}$) over a 8 \AA-thick slab inside the 20NS structure, \textcolor{red}{before (left, $\epsilon=0$) and after (right, $\epsilon=0.4$) fracture. $n_{\rm c}$ is calculated from the local sodium oxide concentration on a square grid of 8 \AA\ in resolution.} The grey area indicates the extent of the final crack. The plot on the upper left corner shows the standard deviation of $n_{\rm c}$ as a function of the \textcolor{red}{grid resolution}.
}
\end{center}	
\end{figure}

Such a fracture, occuring via cavity initiation and coalescence, is direct evidence that the glass should not be treated as a homogeneous material at the nanoscale. Indeed, at this scale, the composition in sodium silicate is inherently non-homogeneous \cite{mauro2013statistics}. The disordered structure of silicate glasses consists of a network of SiO$_4$ tetrahedra forming some rings \cite{Marians1990}. Network modifiers cations, such as sodium, depolymerize the Si--O network and thereby increase the average ring size \cite{Du2004, Du2005}. Studies focusing on the medium range structure of silicate glasses have identified as large as 20-member rings in 20NS \cite{Du2004}, which give rise to spatial fluctuations of composition \cite{Bauchy2013, Du2004}. On the other hand, it has been established that, for (Na$_2$O)$_x$(SiO$_2$)$_{1-x}$ glasses, the rigidity of the structure, as indicated by the number of constraints per atom ($n_{\rm c}$), directly depends on the composition \cite{Mauro2011, Bauchy2011, Thorpe1983, Vaills2005}:
\begin{equation}
\label{eq:nc}
n_{\rm c} = \left( 11 - 10x \right )/3
\end{equation} As a result, the inhomogeneity in the local fraction of sodium oxide $x$ induces some variations in the local structural rigidity. Such heterogeneity is demonstrated by the $n_{\rm c}$ contour map in Fig. \ref{fig:map}. We can see that, for 20NS, substantial spatial variations of the structural rigidity exist within the glass, when the spatial resolution is kept below 15 \AA.

As described by Eq. \ref{eq:nc}, sodium silicate glass goes through a rigidity transition at sodium oxide concentration of 20\% \cite{Bauchy2011, Thorpe1983, Vaills2005}. At lower sodium oxide concentration, the structure is stressed-rigid and has limited ability to rearrange and relax. On the contrary, above 20\% sodium oxide, in the flexible regime, some floppy modes of deformation are available for the atomic structure to rearrange. Additionally, the existence of an isostatic intermediate phase has been reported for sodium oxide concentrations between 18\% and 23\% \cite{Vaills2005, Micoulaut2008}, corresponding to a theoretical $n_{\rm c}$ between 3.07 and 2.90. As such, we find that, although the 20NS composition should be isostatic, on average, at the macroscale, the compositional variations at the nanoscale result in the formation of flexible and stress-rigid regions in the glass, as illustrated in Fig. \ref{fig:map}. Since the flexible regions feature a lower structural rigidity, they should undergo noticeable relaxation under strong stress, as experienced during the fracture.

A comparison of the atomic structures before and after fracture supports such conclusion. Indeed, during the fracture, cavities preferentially form in the flexible regions and, eventually, lead to a preferred crack propagation through these regions (see Fig. \ref{fig:map}). \textcolor{red}{This is in agreement with experimental observations \cite{Almeida2014}.} In addition, the fluctuations of the local composition also result in heterogeneity of bonding, as Na--O bonds are much more ionic and weaker than the Si--O ones. As such, these two kinds of bonds behave drastically differently under strain. As shown in the inset of Fig. \ref{fig:ss}, as the strain increases, the relative deformation of Si--O bonds presents the same shape as that of the stress-strain curve, and eventually goes back to its initial zero-stress value. This means that Si--O bonds essentially deform in a reversible elastic fashion under stress. On the other hand, the Na--O bonds behave in a significantly different way. Indeed, the maximum relative elongation of the Na--O bonds is much lower than that of the Si--O bonds, which shows that, after a short elastic regime, these bonds yield at low stress and, thereby, initiate the fracture through some plastic deformations prior to the failure of any Si--O bond. This suggests that the observed nano-ductility mainly arises from Na--O bonds and, therefore, should be very limited or non-existent in pure silica.

In addition to their bond lengths, the connectivity of Si and Na atoms are also affected differently during the fracture. Indeed, most of the Si atoms (> 99.9\%) remain four-fold coordinated after the fracture. Most of them also retain the same O neighbors throughout the fracture process, as only a small fraction (1.5\%) acquire new neighbors, mostly as a result of local relaxations at the fresh surface formed after fracture. On the contrary, around 90\% of the Na atoms switch their oxygen neighbor during the fracture, even though the average coordination number only shows a moderate change (from 5.94 to 5.46). Such exchanges of neighbors are irreversible, which clearly shows that a significant number of plastic deformations happen around Na atoms. It is also worth noting that the local relaxation around Na atoms can happen at stresses much lower than the strength of the glass. \textcolor{red}{This feature echoes with the flexible nature of Na--O polyhedra \cite{Bauchy2011}, and may be related to the observed relaxation of alkali silicate glasses at low temperature \cite{Welch2013, Yu2015}}. Finally, bond-angles are also affected during the fracture. Although strong intra-tetrahedral O--Si--O angles remain largely unaffected, a fraction of the weaker inter-tetrahedral Si--O--Si angles \textcolor{red}{experience a permanent change of their average value after fracture, which suggests that some plastic deformations occur in the Si--O network}.

\begin{figure}
\begin{center}
\includegraphics*[width=\linewidth, keepaspectratio=true, draft=\ddst]{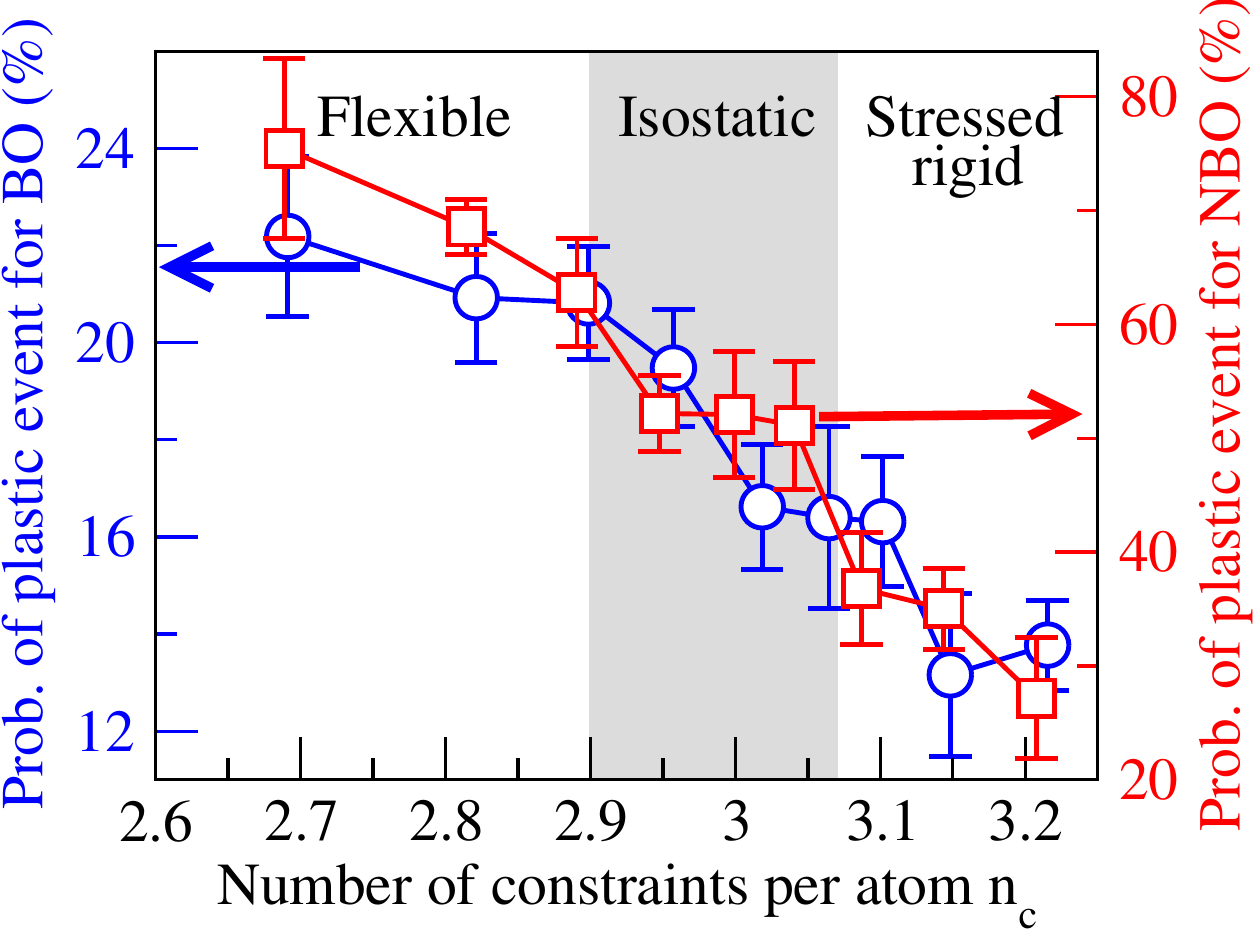}
\caption{\label{fig:plastic} (Color online) Probabilities of plastic events (see text) occurring around the bridging (BO, \textcolor{red}{open circles}) and non-bridging (NBO, \textcolor{red}{open squares}) oxygen atoms with respect to the local number of constraints per atom $n_{\rm c}$. The sampling frequency in terms of $n_{\rm c}$ is chosen so that each group contains at least 500 oxygen atoms. The grey area indicates the boundaries of the intermediate phase observed experimentally through modulated \textcolor{red}{Differential Scanning Calorimetry} \cite{Vaills2005, Micoulaut2008}.
}
\end{center}	
\end{figure}

Based on these observations, we classify and quantify the fracture-induced plastic events happening in the environment of (1) the bridging oxygen (BO) species, i.e., O linked to two Si atoms, and (2) the non-bridging oxygen (NBO) species, i.e., O linked to one Si and at least one Na atoms. (1) The number of plastic events affecting BOs is defined as the number of Si--BO--Si angles that show a significant permanent distortion of at least 15\%. (2) The number of plastic events affecting NBOs is defined as the number of oxygen atoms that lose at least half of their initial Na neighbors. The probabilities of these plastic events are then correlated to the local rigidity experienced around the oxygen species, as obtained by calculating $n_{\rm c}$ within a 8 \AA\  radius sphere centered around the considered O, using Eq. \ref{eq:nc}.

As shown in Fig. \ref{fig:plastic}, both for BOs and NBOs, the probability of plastic events decreases with increasing local rigidity, as captured by $n_{\rm c}$, which shows that ductility is mainly concentrated in the flexible regions. Interestingly, the probability of such events shows a plateau in the stressed-rigid domain. Such a trend appears similar to the fraction of floppy modes observed in chalcogenide glasses \textcolor{red}{\cite{Kamitakahara1991,Mauro2007, Mauro2007a}}, which suggests that plastic events arise from such low energy modes of deformation.

Overall, these results are consistent with the following topological picture. Thanks to internal degrees of freedom, the flexible regions feature plastic deformations. On the contrary, due to the high number of constraints, stressed-rigid regions are locked and unable to reorganize under stress. Eventually, due to the heterogeneity of the local rigidity, plastic events occur in different regions of the glass, and ultimately merge to form the crack, resulting in a nano-ductile fracture. On the contrary, pure silica glass shows very limited heterogeneity and, thereby, breaks in a brittle way through a catastrophic failure of Si--O bonds. This also suggests that heterogeneous multi-component glasses that are isostatic overall should feature the highest amount of nano-ductility. Indeed, flexible glasses feature a large amount of flexible domains, percolating through the bulk structure, which should decrease the probability of crack deflections. On the contrary, stressed-rigid glasses possess a low amount of flexible domains, which limits the number of possible plastic events.

Since the heterogeneity of topological constraints remains limited to a nanometric scale, this nano-ductility is unlikely to result to micro- or macro-ductility \cite{PhysRevLett.114.215501}. Nevertheless, more complex glasses characterized by phase separation or long-range heterogeneity could be considered to maximize this ductility and, thereby, increase the resistance to fracture \cite{mauro2013statistics}. Pressure, as applied during quenching, can also affect heterogeneity and, consequently, nano-ductility. Indeed, pressure has been found to lower the extent of topological heterogeneity in sodium silicate \cite{Bauchy2013}, which results in a more brittle fracture \cite{Bauchy2014}. Pressure is, on the contrary, thought to induce micro-heterogeneity in pure silica \cite{arndt1969anomalous} and, therefore, appears to be a promising degree of freedom to tune the ductility of glasses \cite{yuan2014brittle}.

\begin{acknowledgments}
The authors acknowledge financial support for this research provisioned by the University of California, Los Angeles (UCLA). Access to computational resources was provisioned by the Physics of AmoRphous and Inorganic Solids Laboratory (PARISlab).
\end{acknowledgments}

\end{document}